\documentclass{soups} 
\usepackage{times}
\usepackage{balance}
\usepackage{booktabs} % for making nice top, bottom, mid lines in tables
\usepackage[flushleft]{threeparttable}
\usepackage{multirow, multicol}

\usepackage{enumitem} %% allows specification of the tab size in enumerate, itemize, ... environments 
\usepackage{paralist} %% needed for compact itemize
\newenvironment{myitemize}
    { \begin{compactitem}[\leftmargin = 0pt $\bullet$] }
    { \end{compactitem} }

\usepackage{xcolor}
\usepackage{wrapfig}
\usepackage{graphicx}
\graphicspath{ {images/} }
\usepackage{lipsum}

\usepackage[hyphens]{url}
\usepackage{hyperref}

\usepackage{subcaption}
\usepackage{enumitem} %% allows specification of the tab size in enumerate, itemize, ... environments 

\usepackage[compress]{cite}

\begin{document}
%
% --- Author Metadata here ---

\conferenceinfo{Submission to Computers \& Security}{\today} 
% \conferenceinfo{USENIX Symposium on Usable Privacy and Security (SOUPS)}{2018. \\ August~12--14,~2018, Baltimore,~MD,~USA.}
\CopyrightYear{2018} 

% --- End of Author Metadata ---

\title{How do information security workers use host data?
    \titlenote{{This manuscript has been authored by UT-Battelle, LLC under Contract No. DE-AC05-00OR22725 with the U.S. Department of Energy.  The United States Government retains and the publisher, by accepting the article for publication, acknowledges that the United States Government retains a non-exclusive, paid-up, irrevocable, world-wide license to publish or reproduce the published form of this manuscript, or allow others to do so, for United States Government purposes.  The Department of Energy will provide public access to these results of federally sponsored research in accordance with the DOE Public Access Plan \url{http://energy.gov/downloads/doe-public-access-plan}.}
    }
    \titlenote{
    Declarations
    of interest: none}
    \titlenote{Corresponding author: Robert Bridges}
}
\subtitle{A summary of interviews with security analysts}

\author{
\alignauthor
    Robert A. Bridges, Michael D. Iannacone, John R. Goodall, Justin Beaver\\%, Kerry Long\\
    \affaddr{Cyber \& Data Analytics Division, Oak Ridge National Laboratory, Oak Ridge, TN}%\\
    %\affaddr{{Intelligence Advanced Research Projects Activity, Office of the Director of National Intelligence, Washington, DC}}
    \email{\{bridgesra, iannaconemd, jgoodall, beaverjm\}@ornl.gov}%, kerry.long@iarpa.gov}
}

\maketitle

\begin{abstract}
Modern security operations centers (SOCs) employ a variety of tools for intrusion detection, prevention, and widespread log aggregation and analysis. 
While research efforts are quickly proposing novel algorithms and technologies for cyber security,  access to actual security personnel, their data, and their problems are necessarily limited by security concerns and time constraints. 
To help bridge the gap between researchers and security centers, this paper reports results of semi-structured interviews of 13 professionals from five different SOCs including at least one large academic,  research, and government organization. 
The interviews focused on the current practices and future desires of SOC operators about host-based data collection capabilities, what is learned from the data, what tools are used,  and how tools are evaluated. 
Questions and the responses are organized and reported by topic.  
Then broader themes are discussed. 
Forest-level takeaways from the interviews center on problems stemming from size of data, correlation of heterogeneous but related data sources, signal-to-noise ratio of data, and analysts' time. 
\end{abstract}

\section{Introduction}
\label{sec:intro}
Networked computing assets have become integral tools for personal, business, and state functions, and most enterprises rely on an operational team of security experts working regularly to ensure networks are healthy and secure. 
Modern cyber security operations now have a diverse collection of tools for  monitoring devices and behaviors,  blocking intrusions, alerting on known problems, and logging a large and increasing set of data. 
In parallel, research efforts, especially in data science and related fields, are proposing and testing new algorithms, technologies, and interfaces to address information security needs.  
Yet, the two communities\textemdash cyber security operations and researchers\textemdash inevitably develop separately. 
Researchers often do not have regular interaction with operators, and cyber operations are generally sensitive about sharing information to limit their vulnerabilities.

To bridge this gap, we have interviewed security analysts to learn more about security operations work practices and tools. 
Our focus was on how operations use host-based data.  
This is for a number of reasons. 
First, host data can give much greater insight into workstations and servers than network data. 
Second, increasingly network data is encrypted forcing security operators to rely on host-based security. 
% is at the cost of greater I/O (input-output, referring to data transfer between multiple systems), processing, and storage costs; hence, as technology develops, greater utilization of this data is becoming possible. 
Third, the  amount of data collected and processed by cyber operations is large and increasing, so continuing to distribute the burden to nodes is a promising avenue for expanding security measures.  
Finally, we foresee an increase in the utilization of cloud resources, where monitoring is possible from both the hypervisor and inside the host but network monitoring is not possible. 
The research presented here complements our work to survey the research literature on host-based intrusion detection~\cite{glass2018survey}.

In this paper we provide an overview of related studies of security operations (Section \ref{sec:related-works}), discuss our methodology in Section \ref{sec:methodology}, then present the interview questions and summarize and report findings from operators' responses from Section \ref{sec:results}. 
Finally we illustrate broader conclusions and insights in Sections  \ref{sec:tangents} and \ref{sec:conclusion}. 

% \textcolor{blue}{this seems out of place, not sure why it is here}

% For quick reference Table 1 gives links and short descriptions of commercially available tools mentioned by respondents. 

\section{Related Work} 
\label{sec:related-works}
User studies to understand the interplay of sociology and technology with respect to security is an active area of research, especially in the Usenix community, e.g., \cite{mcgregor2015investigating, mcgregor2017when}.  
Here we give a brief survey of those related works, defined as user studies to understand SOC operations. 
Compared to research proposing novel algorithms and tools for SOC operations, e.g., \cite{best2010real, bridges2018ideas,  chen2017automated, hossain18dependence, huffer2017sansr, glass2018survey, goodall2018situ, ma2016protracer,   stucco}, there are relatively few works engaging the actual security workers to inform these efforts. 

Early research of Goodall et al. \cite{goodall2004work} focuses on intrusion detection tasks conducting contextual interviews in situ with nine security analysts. 
The authors describe a generalized procedure, i.e. the steps analysts follow for handling alerts from monitoring to incident analysis to response. 
Primary findings are that domain expertise (understanding attacks, protocols, etc.) are needed in addition to tacit, network-specific knowledge and that challenge of contextualization by manually integrating data from multiple sources is noted. 
Two primary suggestions based on the research findings are proposed by this work:  (1) a tiered structure to SOC  analyst roles and (2) visualization tools to gain efficiency in each portion of the process.

Fink et al. \cite{fink2005visual, fink2006bridging} conduct semi-structured interviews with 20 security workers focusing on the manual process to mentally correlate network- and host-level data. 
The work describes a new technological solution to address the problem, HoNe (Host-Network Visualizer) that correlates host processes with corresponding ports in an intuitive visualization. 

Fourteen security professionals from five organizations participated in a semi-structured interviewed study by Botta et al. \cite{botta2007towards}, who seek information about the roles, responsibilities, tasks and skills of SOC employees. 
Their findings are that the most important skills of security analysts are pattern recognition, inference, and ``bricolage'', the ability to build a story from diverse ingredients. 
They also provide a table of tasks mapped to corresponding tools.

Werlinger et al. \cite{werlinger2009integrated} conduct 36 semi-structured interviews with security management and professionals from 17 organizations with the goal of itemizing and understanding relationships between challenges faced by security experts. 
Many of the 17 challenges Werlinger et al. identifies are supported by our work, especially, lack of training, lack of budget, priority of security within the organization, management, vulnerabilities, and lack of effective tools. 

Follow-on work by an overlapping set of authors, Werlinger et al. \cite{werlinger2010preparation} interview 16 practitioners with an eye on incident response practices reaching similar conclusions to Goodall et al. \cite{goodall2004work}---that tacit environmental knowledge was key. 
They also discuss the collaborative nature of the work requiring discovery of what happened from tools/data well enough to articulate it to other parties. 

Bauer et al. \cite{bauer2009real} focus on access-control management, interviewing 13 operators with expertise in access control of data systems or physical space. 
They cite disconnections between capabilities and policy and between policy makers and implementers as the primary hurdles in the space. 

A limitation of the works above, as well as the work in this paper, is the small sample size that may limit the generalizability of the results.  
This is a product of the security operations' default to not discuss details with external or untrusted parties as well as their work volume.   
De Souza et al. \cite{de2011information} avoid this limitation by supplementing a qualitative study of 20 semi-structured interviews with a quantitative study, delivering an electronic, 45-question survey and receiving over 200 respondents. 
De Souza et al. focus on the information needs of the system administrators, and the study was conducted on a particular ``Big Service Factor'' (BSF) that provides IT (information technology) services to many client enterprises; i.e., they study IT security professionals that are outsourced, rather than IT staff working exclusively internal to the networks they protect. 
In parallel to findings of Goodall et al. \cite{goodall2004work}  and Werlinger et al. \cite{werlinger2010preparation}, De Souza et al. reconfirm that network-specific knowledge is a key skill.
De Souza et al. conclude that interaction with the  customer and knowledge of the customer's network are the key components for BSF security personnel. 

Sundaramurthy et al. \cite{sundaramurthy2016turning} employ anthropological methods to study SOCs by shadowing and working in the specific roles at the SOCs they study. 
This gives unique perspectives from which to build tools. 
Their findings focus on conflicts of interest, or ``contradictions'' created by the environment, job, management, and tools. 
For example, analysts describe being driven by continually learning to keep up with  evolving threats, yet have ongoing pressure to prove their worth to management through metrics, reports, etc. 
Many of the contradictions  mirror our findings and are noted below when discussed in our our results. 
Sundaramurthy et al. also describe a  tool that was fielded at a SOC because it automated repetitive tasks, but their studies found that without continual development to evolve the tool with the operation, it was quickly abandoned. 
This unique study provides a variety of challenges imposed on the security professionals as well as actual experience of challenges for researchers attempting to develop adopted tools. 

To our knowledge this is the only work that interviews SOC operators to empirically study how they use and evaluate host-based data and related tools.

\section{Methodology \& Participants} 
\label{sec:methodology}

Our interview consisted of 25 questions, with one question on the interviewee's professional role/title and the remaining 24 broken into categories/sections of this paper  as follows: 
\begin{myitemize}
\item {\it Questions on host-based data, Section \ref{sec:host-data}:} These questions probe what and how much data operations collect and how it is stored and finally what data is desired by the participants. 
\item {\it Questions on tools, Section \ref{sec:tools}:} This section reports the commercial tools adopted by operations, their turnover, and how internal tool development occurs. 
\item {\it Dynamic collection capabilities, Section \ref{sec:dynamic}:} This section covers questions pertaining to just-in-time or otherwise automated changes to data collection to enhance security in certain circumstances. 
\item {\it Learning from host-based data, Section \ref{sec:learning}:} These questions ask what information is desired/obtained from the data collected. 
\item {\it Evaluation of Tools, Section \ref{sec:eval}:} In this section we report on how operations evaluate the security tools used, and specifically probe how analysts' time and end-user costs are take into account. 
\end{myitemize}

We interviewed 13 respondents from five distinct organizations including at least one academic, one government, and one large research institution. 
Some respondents made clear that their organization did not own the hosts on their network, which limits their host-based control and data collection capabilities and incentivizes network-based means. 
Conversely, other participants worked for organizations that did own their hosts facilitating enterprise-wide, host-based security measures. 
Of the 13  participants, five responded with a role of Security Analyst or equivalent, three were Senior Security Analysts, one was a Chief of Host Forensics \& Incident Response, one was a SOC manager, one was a manager of both a NOC and SOC, one was a Network Engineer, and one was the Client Services Team Lead. 

The interviews were semi-structured, following the set questions, but allowing analysts to digress. 
Interviewers recorded answers for future analysis. 
The interviews were conducted one-on-one, five over the phone and the remaining eight in person with duration of approximately an hour on average.

We follow standard qualitative research methods~\cite{strauss1990basics}\textemdash  reading responses and letting themes surface while trying to be unbiased, objective, and not presuppose a result. 
We note that some amount of technical ``pre-knowledge" in cyber security is required to understand the domain and interact with participants; so while we did not enter the coding phase with themes in mind, we did have a working understanding of the domain of cyber security.
More specifically, our workflow for coding results was to shuffle responses, read all responses for a set of related questions, then report on our findings. 
For questions requiring quantifiable or categorical responses, we report aggregate statistics. 
When more open-ended responses were common, we highlight representative responses and attempt to accurately summarize the findings. 
Care was given to report useful results while not disclosing details of any organization or person. 
Finally, vendor-specific details, when encountered have been anonymized.

Our experience is that a relationship of trust in our research team by someone at the institution was seemingly a requirement for anyone to agree to an interview. 
We suspect that this is in part the nature of security operation centers---to be reluctant to share information about themselves with un-trusted parties and to protect their time. 
It was clear from all interviews that the operators are extremely busy, and their time is very valuable---a theme also borne out in the needs and desires for analytic tools.

\section{Questions \& Results} 
\label{sec:results}
Our interview questions were organized into groups related to: host data collection, tools used in operations, dynamic data collection, learning from data, and how tools are evaluated. This section presents each of the interview questions and reports the results for each. 

    \subsection{Host Data Collection Questions} 

\label{sec:host-data}
The first section of the interview dealt with the collection, size, and storage of host-based data, as well a discussion of data analysts do not currently collect that could be helpful. 

{\it Questions 1 \& 2: What host-based data do you collect? Are there any special data collection methods for virtual hosts?: } All interviewees' organizations included host-based anti-virus  (AV) tools and many reported collecting logs of alerts from these tools by default. 
Low-quantity logging was expressed by employees of a few  organizations, e.g., only performing periodic scans to identify if personally identifiable information (PII) was stored on workstations, logging failed authentications, and collecting host-based AV logs. 
These institutions often also collected more thorough logs from servers. 

Many organizations had ample collection capabilities and by default store a wide variety of host event logs. 
Logs from end-point AV, authorizations, system and accesses, security system alerts, Windows Registries, DLLs, VPN (virtual private network) remote connections, and vulnerability scans are examples mentioned.   
This capability varied widely  (a) for servers versus workstations with more logs collected from servers in general and (b) across operating systems (OSes) with more logs for Windows workstations than Linux and Mac OSes. 
It was not uncommon for these event logs to be collected for all workstations and servers by default, while others mentioned it was too costly for all workstations. 
Some organizations collect virtual private network logs, e.g., IPs, MAC addresses, authentication information. 
Scanners that report if PII  occur on host data are used. 
Email information and user information is commonly available to the SOC for each workstation. 
Management systems that enforce policies and allow automated configuration of data collection settings on the corresponding host clients are used. 
No operators reported augmented data collection for virtual hosts on their network. 
Often the ability to actively obtain current host data (e.g., hard drive (HD) images) that is not passively collected was communicated.

{\it Question 3: How big is it?: } Reported size of host data varied widely among those that responded to this question. 
On the low end an approximation of 300MB/day were given. 
One respondent works across many organizations and reported 100GB to 10TB per day, with the latter the largest estimate given during our interviews. 

% {\it Question 3: Storage \& Query: } 
%Spunk, %(\url{www.splunk.com}), 
% a commercial event,  logging, and management tool was mentioned by many participants and appears to safely be the primary tool for indexing data. 
% Two other tools were mentioned for indexing logs, Elasticsearch, % (\url{https://www.elastic.co/}), 
% an open source database, and Snare,  %(\url{www.snaresolutions.com}) 
% another open source  event management tool, yet, both these operators mentioned Splunk in use as well. 
% Overall, Splunk 
{\it Question 4: How long do you store it for analysis?: } 
Commercial event logging and management subscription costs were cited directly by some as the constraint for data collection after mentioning they would benefit from more data collection.  
Perhaps this is unsurprising given estimates from the numbers above---the sheer quantity of host data collected and available to security operator centers is between 1GB-1TB/day, stored for 3 months$\approx 100$days = 100GB-100TB. 

Some participants mentioned open-source logging and management software. Specifically, a few said both commercial and open-source solutions were in use, to either save costs, or test and potentially transition to a new solution. 

For storage duration, three to six months was standard. 
Many reported using or having future plans of a three-tiered scheme\textemdash moving data from the initial database after three to six months to a less queriable but cheaper storage mode for two years, and finally to tape (presumably cheaper and less liquid) for six years or ``indefinitely''. 
Forward leaning operation centers are targeting a data-dependent pyramid scheme to cut costs and increase efficiency of queries. 

{\it Question 5: What host-based data is not currently used for detection that you think could be really beneficial for detection?: }
To our question on desired host-data collection capabilities we received many diverse responses illustrating the current limitations (or lack thereof) and uses of host data. 
Two respondents indicated their current host-data collections were sufficient, while others desired capabilities to collect more data from hosts, in particular application-level audit logs.  
Greater logging of host activities on non-Windows machines was requested. 

Two participants indicated a desire for tools allowing visibility of packets before encryption and transmission. 
Interestingly, one participant requested something akin to a remote desktop recording allowing analysts to see what the user was clicking/typing at the time of a past incident. 
Unsurprisingly, increasingly used encrypted traffic is forcing analysts to leverage host-data and require novel tools to balance security with privacy. 
This is also articulated by Hossain et al. \cite{hossain18dependence}.

Multiple respondents asked for contextual integration of data. 
One requested automated matching of file hashes to corresponding processes and two others sought correlations related to the host's ports, including which are listening, in use, and correlating ports to the processes using them.  
Perhaps non-coincidentally, interviews of Fink et al.~\cite{fink2005visual, fink2006bridging} of security analysts identified exactly this problem\textemdash generally, related but heterogeneous data from different systems must be held in the analysts mind and manually correlated, and, specifically, process information (host level) is problematically manually correlated with corresponding ports, packet transfer, etc. (network information) by analysts. 
Fink et al. \cite{fink2006bridging} introduces HoNe, a visual analytic tool to address correlating heterogeneous but related logging data for cyber analysts. 
This illustrates the gap between research-developed cyber tools and those adopted by analysts or through industry-licensed products.

While most operators collect end-point AV logs from workstations, analysts requested a technology that furnishes privilege escalation notifications to the user. 
They explained that they regard the user as the authority on what should happen on their platform.

    \subsection{Tools Questions}
\label{sec:tools}
This section summarizes responses to questions probing what tools are used by operations for host-based data collection. 
% See Table~\ref{tab:tools} for products mentioned in use by interviewees. 
Additionally, we asked how new tools or data feeds are integrated. 
We included questions on what and how in-house, situational, or  one-off developments of code or tools take place, in hopes of stimulating discussions of how new tools are developed. 
Finally, we asked analysts to identify their oldest, newest, and most critical tools. 

\begin{figure}[h]
\centering
\includegraphics[width = \linewidth]{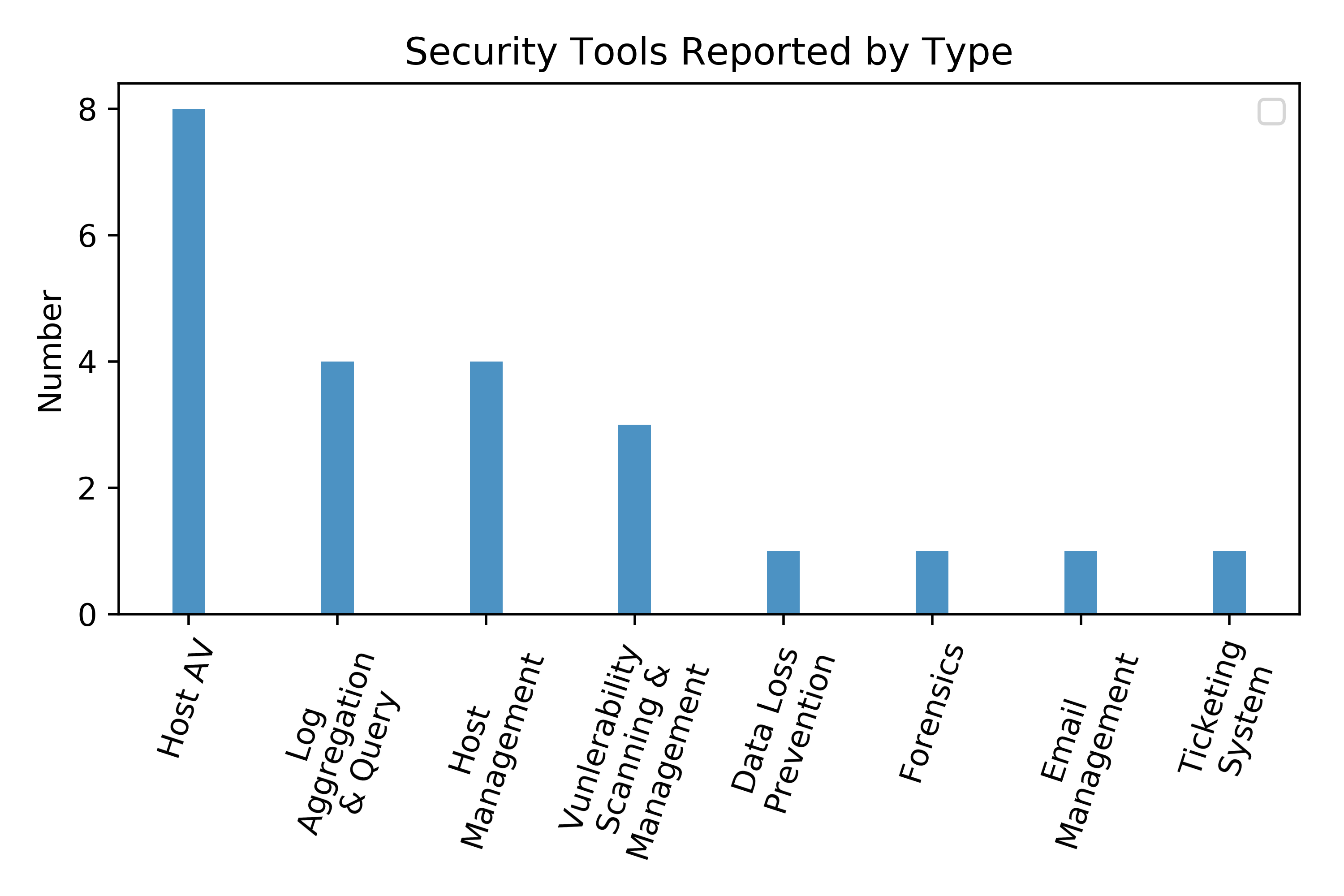}
\caption{\small{Bar chart depicts participants responses to ``What detection products are used at the host level?''. Responses are binned by category.}}
\label{fig:tools}
\end{figure}

\textit{Question 6: What detection products are used at the host level?: } Answers to this question were aggregated by type to anonymize the vendor or product specifics and are presented in Figure \ref{fig:tools}. We note that some tools, may experience low frequency because whether they count as ``host level'' is debatable, e.g., ticketing systems. 

\textit{Question 7: Can you describe tools or code developed in-house for detection or investigation?:}
While many respondents reported no custom tools were in place, others reported a surprisingly robust set of  hand-crafted tools. The reported custom tools are as follows: 
\begin{myitemize}
\item Python, bash, and shell scripts for blocking, filtering based on regular expressions, and parsing;
\item Work to develop signatures and perform dynamic malware analysis  tools (i.e., tools for analyzing malware while running in a virtual environment);
\item Scripts for identifying port scanning and reverse-DNS queries  that block detected traffic at the firewall; 
\item Code to identify then log and block password brute-force attempts;
\item Configuring dashboards with SIEM tools for monitoring authentication data and other data of interest;
\item Vulnerability management tools, which correlate information from many other systems;  
\item Client services that perform work on the host, then report results;
\item PII-identifying scanners were crafted but have now been replaced by [a commercially available tool];
\item Unix Logging Service (ULS), i.e., a Unix analogue for collecting Windows event logs that allows configurable log forwarding from Unix-based hosts to the operation center; 
\item Network flow visualizations by port; 
\item Customized [end-point configuration management tool] %(\url{www.ibm.com/security/endpoint-security/bigfix}, 
 scripts for stopping malware propagation that AV cannot detect. 
\end{myitemize}
Finally, some analysts have piloting and adopted tools from professional research projects. 

\textit{Question 8: Can you discuss the process for deciding to integrate new data feeds or  security tools (especially host-based detection tools)?: } 
Responses showed a similar workflow from most participants' organizations, although criteria in each step varied. 

Motivations for considering on-boarding new tools or feeds varied from technological advances, ending life-cycle of current tools, changes in policy or law, to appease management, or to fulfill needs identified by analysts. 
Most discussed a period of investigating what tools/feeds were available in comparison to what is currently in use with multiple respondents emphasizing complementary tools. 
One analyst consulted a visualization created to display the present tools and feeds. 
Another mentioned new tools must be ``disruptive or novel'' to be adopted. 
Overall, this phases of decision-making seemed to focus on whether the desired tool complemented current tools, met a need, and produced understandable output; furthermore, it was often a result of managerial- or policy-prescribed changes.  

The next phase of the process described by most respondents involved some evaluation of the tools  ease-of-use and security posture itself. 
Some consulted a  list of approved/unapproved objects, or looked for how well-adopted the tools is. 
Others mentioned the usefulness of the output was a large factor; specifically, coarse, untrusted, and/or not-understandable output (e.g., a red/yellow/green security rating) was commonplace but not helpful, and overly abundant although precise information (e.g., a multi-page, detailed vulnerability report) was similarly not useful.  
Analysts' time was a large factor, citing a tools ease of install, configuration, and use as deciding factors. 
In particular, one participant mentioned the default was to purchase install/configuration services from the vendor alongside the product/license. 
We (the authors) investigated the issue of install and configuration support for six commercially available security products, all of whom now provide seemingly ample (e.g., 160 man hours) initial support. 

Other concerns voiced were security vulnerabilities imposed by the tool under consideration, e.g., if the tool requires high user privileges to run. 
This respondent operated under the assumption that ``vendors do not [penetration] test their own tools''. 

One analyst explained that testing tools in advance is worthwhile to reveal that they are slow regardless of the resources of that particular machine. 
In general, cloud-based tools are ``not responsive''. 

The third stage of evaluation of new tools/feeds involved hands-on interaction. 
One analyst mentioned his organization has a ``bake off'' to test similar tools head to head  before deciding, while others discussed at least a trial period of a tool. 
One organization has a tier schema for testing such tools with SOC testing them first, IT (NOC) second, and then a cost-to-user evaluation (presumably based on the performance in the tests). 

Finally, two respondents mentioned upon final decision to introduce a new tool/feed, a plan is implemented to roll out the tool to different user groups in phases. 

{\it Questions 9, 10: Can you describe how/when you write one-off code for detection or investigation, and how do one-off efforts develop into new detection or investigation tools?: } 
Four respondents said this practice either never happened or was very rare. 
Another said it is commonplace whenever something is found in the wild that poses a threat to more than a single host. 
Examples included writing detection scripts for malware that was spreading but undetected by their standard host AV. 
Other examples provided by respondents included augmenting configuration of current tools, e.g., developing and pushing new signatures to an intrusion detection/prevention system (IDS/IPS), developing new queries in log management software to be used for specific investigations,  and customizing host configurations via one-off scripts. 
Other examples included pre-patching vulnerabilities manually, developing Python scripts to analyze encrypted packets, and searching host files through scans or specific disk images. 

The majority of participants declined response to how the single-instance tools develop into customized tools. 
One pointed out that this is rare simply because one-off code is by definition for a very specific instance. 
Two mentioned that the inadequacy of their custom tools prompted purchase of commercial versions, specifically citing log management and query tools and PII scanning tools as examples. 
Others pointed out that most one-off code is built on or leverages their existing tool suite. 
As such it becomes adopted through  configuration changes. 
For those that responded positively, they pointed to continual development driven by need until it was production quality. 
A few mentioned handing ideas to professional researchers and iterating with the researchers with suggestions and evaluations during development stages. 

\begin{figure}[h]
\centering
\includegraphics[width = \linewidth]{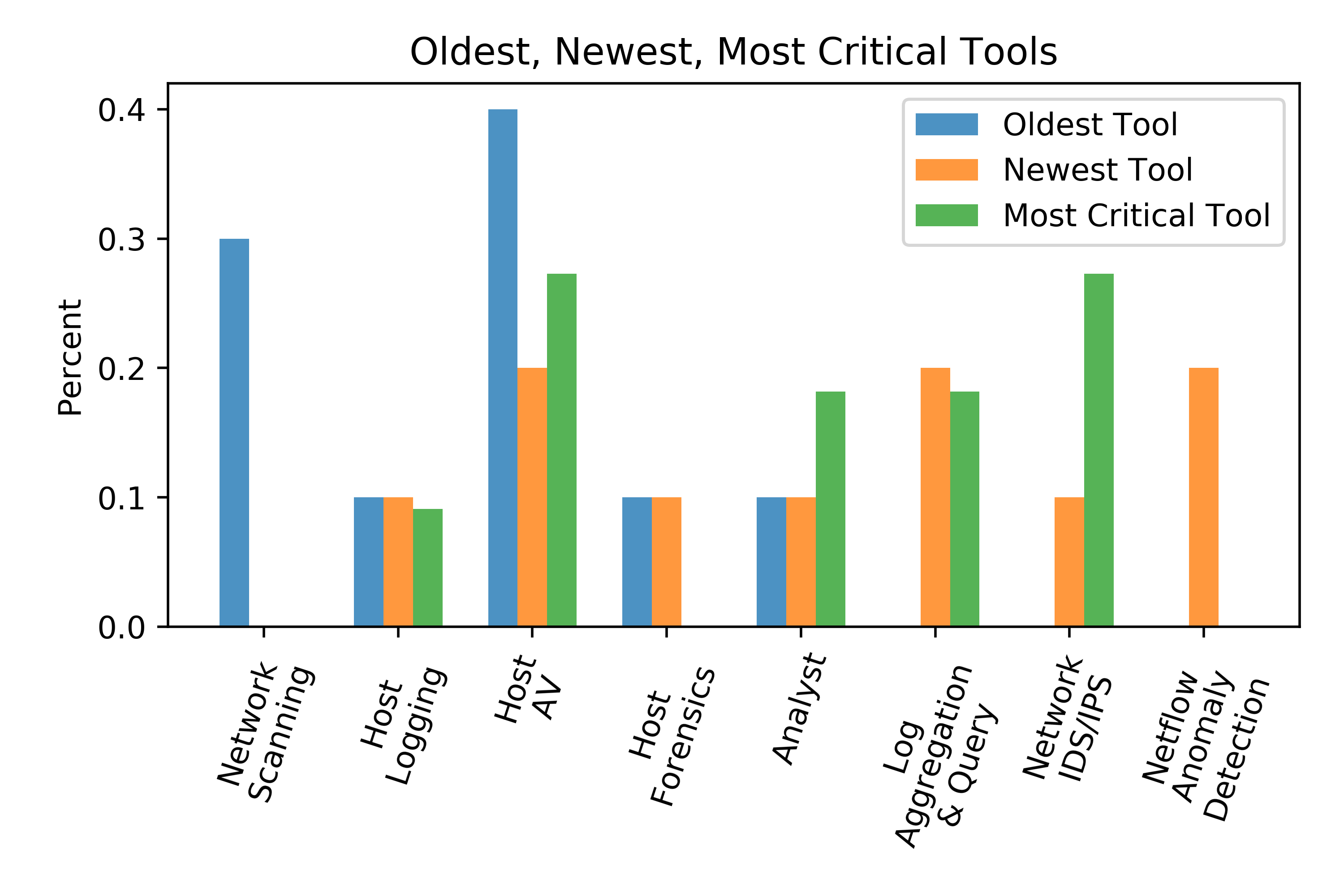}
\caption{\small{Bar chart depicts participants responses to ``What tool is the oldest? Newest? Most critical?''. Responses are binned by category. Network Scanning category includes tools that identify assets, vulnerabilities, or data on the network, e.g., Nessus. Host Logging category refers to configuration software for what system logs are collected. The Analyst category refers to the security personnel themselves.}}
\label{fig:old-new-mc-tools}
\end{figure}

\textit{Question 11: What is your oldest detection tool? newest? most critical?: }
Finally, we asked analysts to name their oldest, newest, and most critical security tools. 
Results are displayed in Figure \ref{fig:old-new-mc-tools}. 
Host AV and network-level IDS/IPS were the top scoring most critical tools. 
Interestingly, Host AV is  the highest scoring in all categories, indicating it is experiencing high turnover, but has been a mainstay for security. 
We note that many analysts without prompting replied ``The Analyst" to these questions, referring to the actual security worker. 
Log Aggregation \& Query tools, e.g., Splunk, were a commonly cited new tool indicating that some operations are only recently adopting software to collect and manage their logs. 
We note that network flow anomaly detection systems are becoming adopted.

    \subsection{Dynamic Collection Questions} 
\label{sec:dynamic} 
Anticipating that the quantity and the signal-to-noise ratio of data collected in operations are problematic, we hypothesized that operations may have or desire just-in-time security tools.
That is, automated changes to the level of monitoring to maximize the security posture and minimize the cost of data collection. 
An example may include triggers to increase data collection when it seems most necessary, rather than collecting such large quantities all the time. 
We included the following questions on current and desired dynamic collection: 
\begin{myitemize}
    \item Do you have the ability to automatically perform dynamic data collection, e.g., to turn on/off some collection capability in certain situations? 
    \item Do you have dynamic or just-in-time prevention or investigation measure that are applied when an intrusion is detected, and what would you like developed? 
    \item If you had the capability to dynamically collect high-fidelity data, how would it be most helpful? 
\end{myitemize}

{\it Question 12, 13: Do you have the ability to automatically perform dynamic data collection, e.g., turn on/off some collection capability in certain situations? \& Do you have dynamic or just-in-time prevention or investigation measure that are applied when an intrusion is detected, and what would you like developed?: }
Multiple responses were negative or unsure about current dynamic collection capabilities. 
On the contrary, many analysts pointed out standard capabilities, such as, AV quarantining a file, blocking actions upon a signature firing (e.g., black-/white-listing IPs), and adjusting parameters in configuration consoles allowing specification of data collection per user, location, etc. 
Some analysts cited abilities to write rules in configuration management software which conceivable could be used to grab more fine-grained data upon certain triggers. 

Many also mentioned that they had capabilities to collect extra, high-fidelity data or change settings of hosts in certain circumstances but the process was partially manual. 
Specific examples included using tools to acquire a disk or memory image of a host under investigation,  changing configuration settings per host at times, shutting down power, and black-listing executables.

{\it Question 14: If you had the capability to dynamically collect high-fidelity data, how would it be most helpful?: }
Many operators generally knew what extra data they would like, but were unsure of the appropriate trigger for automatically collecting it. 
Responses included memory dumps, disk images, and recording user actions at critical times. 
Automated network segmentation was requested to strengthen the security posture dynamically. 
At the host level download or execution of files were requested triggers for elevated data captures. 
Two separate operators described network level events as desired triggers for such capabilities, mentioning white-/black-lists and anomaly detectors. 
Both requested automatic interrogation of the host to   correlate the file hashes and processes bound to network activities that triggered an alert.

Operators often cited the need for integration of information from different tools and event correlation. 
One operator requested a tool to automatically correlate user and system data (e.g., user's ID, email, and organizational role along with that host's OS) across multiple internal systems. 
Multiple users explained their workflow included manually correlating events across systems to fill out an incident ticket, and they desired automated population of tickets across different data logging systems. 
Specific requests included a ticket generator that automatically populates alert info along with full packet captures from a time window surrounding the alert and, separately, some capability to make correlations across multiple hosts. 
One analyst relayed  an example\textemdash that SQL attack alerts require him to manually pull and look into packets to determine if it is a true or false alert, but if the alerting tool could include the packets with a ticket it  would save four hours a week.

Instead of dynamic, in-depth data collection, one respondent preferred to collect all possible host data and have a way to ``distill'' data and prioritize what to store. 
Another believed enough data was collected by his organization and the need was better filtering, access, and visualization of the data. 
Worries of dynamic collection capabilities producing overwhelming quantities of data  were expressed. 
Multiple participants requested greater accuracy of detection tools, and voiced concerns of false positives limiting the effect of any automated tools. 
More specifically, an operator requested tools that produced more fine-grained categorization of events through integrating with other data streams. 
He explained that some tools alert while others both alert and give a category of the event or malware detected, yet in both cases the categories are broad requiring analysts to  determine if it is a false positive. 

    \subsection{Learning from Data Questions}
\label{sec:learning} 
This portion of the survey asked the analysts what they are or desire discovering from their data, and challenges they face using it. 

\textit{Questions 15, 16: What information are you discovering from data generated at the host? What would you like to discover?: } As the answers given often conflated what information is learned and what is desired, we report both together noting the clear wishes that are not yet available. 
Many answers described a broad set of goals including indicators of compromise, misuse, abuse, and misconfigurations\textemdash as one analysts put it, ``a narrative of what happened on the network and [to] dissect what happened''. 
Others seemed focused on data exfiltration stating their goal is to find what data, by whom (attribution), and the how it moved (the sequence of events). 
Specific mappings from data types to discovery/desired discovery were given: end point AV scans were reported as host data used to diagnose the hosts' health; new hashes and permission changes help identify suspicious files; running processes, open sockets, and objects in memory are used to search for suspicious connections; social security and credit card numbers are identified to locate hosts housing unauthorized PII; failure reports e.g., bluescreen errors, are investigated to identify software bugs and patterns of failure; authentication errors and privilege elevations are triggers to investigate unauthorized accesses. 
Desires in this vein included mobile device heath monitoring, session locks, powering on/off of hosts, command line inputs, and file actions such as renaming, copying, moving commands.

Many responses indicated host data analysis required integration of information across various sources and over time with the requests for tools providing general help in this regard.  
Specific requests included discovery of 
``what application-level logs mean'', 
and correlation of network statistics of listening connections with process, port, time, and the IPs to which they are talking. 
This goes hand-in-hand with the problematic signal-to-noise ratio which was also explicitly identified as a problem. 
One respondent asked for tools to filter non-useful data, reduce volume overall, and focus on important logs. 
He/she gave specific examples of desired information, such as, terminal sessions interacting outside the network (shells that spawn a socket) and pointed to Microsoft Sysmon %\footnote{\url{https://docs.microsoft.com/en-us/sysinternals/downloads/sysmon}} 
as an ideal example\textemdash Sysmon allows  configuration to collect a subset of host logs most relevant to security investigations, e.g., process creations linked to creator processes/commands, DLLs, disk read/writes, file hashes, and information related to creation and modifications to name a few.

    \subsection{Evaluation of Tools Questions} 
\label{sec:eval}
We suspect that purchase and continued use of a tool follows a mental, although possibly not explicit calculation--- one must balance a tool's contribution to security against the variety of costs, e.g., imposition to end user, analysts time for configuration, reconfiguration, and false positive investigations, and monetary costs. 
To this end, our final subset of questions probes how security operations evaluate their detection and prevention tools, what the costs of security measures are, and how to reduce costs. 
We explicitly ask about how to measure what the costs are to the user's workstation and the security operators. 

{\it Question 17: How do you measure effectiveness of detection and prevention capabilities and novel tools?: } 
While one respondent reported evaluation did not apply to his/her job, the remaining 12 responses  can be divided into three overlapping viewpoints on evaluation criteria: (a) accuracy of the tool, (b) the cost to the security operation, and (c) the cost to the end user. 
A second dichotomy was evident as well---whether tools were evaluated before or during widespread deployment. 

Most commonly,  the quantity of false positives commonly mentioned with two analysts mentioning the balance of true-to-false positives. 
Some  gave less precise comments such as ``tools are evaluated on whether we can depend on them,'' indicating accuracy of the alerts or data produced is the criteria. 
Others mentioned false positive analysis with comments such as ``[false/true positives/negatives] are taken into account'' or ``are hard to quantify'', suggesting there was no formal evaluation, but it is in their minds when considering the worth of a tool.  
One respondent mentioned false negatives without prompting\textemdash see next question. 

Three different analysts brought up the following false-positive scenario\textemdash a tool alerts or indicates a vulnerability in a host for which  the indicated vulnerability is non-existent or even impossible given the host's configuration. 
One went on to say 
\begin{quote}
90\% of the workload of the SOC could be eliminated if alerts for machines that don't have the needed vulnerabilities were automatically discarded.

- Participant 3
\end{quote}
This supports findings of Sundaramurthy et al. \cite{sundaramurthy2016turning} who reported that analysts spend ten minutes on average to populate tickets with data and repeat this process over 15 times per day, reducing a potentially interesting job to a repetitive mindless task causing burnout.

To perform accuracy evaluations, most respondents alluded to the quantity of false positives found during actual use.
Specifically, that an ensuing investigation spawned by an alert lead analysts to find the alert was errant, and many also said that corroborating results of multiple tools illuminates  false positive or negative when discrepancies are present. 
Multiple different respondents remarked that this led to reconfiguration costs, in particular, re-crafting signatures, or in some cases working with vendors to patch tools.  
Multiple other analysts dove into pre-adoption evaluation measures including testing tools in a pilot environment, with a pilot user subgroup, in sandboxes, or performance (interpreted as computational cost) tests. 
One analysts discussed a mini-research project to test endpoint AV tools head-to-head before widespread use by setting up virtual environments, infecting with a collection of malware samples, and recording accuracy metrics for different tools. 

These pilot testing scenarios bled the conversation from accuracy evaluations to cost of the tools to the security operation. 
As an example of this, an analyst remarked that while ElasticSearch and Splunk are used redundantly in their operation, ElasticSearch is his/her preference because the query time is much faster, results are more precise, and it is more user friendly. 
One respondent outlined a workflow for tool evaluation pre-purchase as follows: 
\begin{myitemize}
    \item Plausibility of deployment is evaluated including the effort to stand up the technology, quality of documentation, ease of install, and considerations of opinions of other adoptees; 
    \item A Pre-deployment trial is conducted with a focus on how much effort is required to obtain actionable information from the tool;
    \item Finally, cost of the tool purchase/subscription is considered. 
\end{myitemize}
The theme of analysts-focused costs were reverberated by others independent of the pre-/post-deployment evaluation time frame, including mentions of the speed of information retrieval (this concern was particularly pointed by one analyst at the tool's performance while by another at the learning curve for use of a tool, e.g., having to learn a query language before use), balancing analysts' time with versus without the tool in place, time used to debunk or verify alerts, time spent on configuration, and time needed by analysts to learn to use the tools. 

Although a minority, a few analysts mentioned impact on the end user, meaning the person working on the endpoint being protected as a consideration. 
Most notably, one respondent seemed primarily end-user focused responding that too much overhead to the host results in the user not actually using the desired capability. 
This response also indicated pilot tests with specific end users. 

{\it Question 18: Does this include accommodations for for false negatives? True negatives?: }
Upon asking, most participants agreed that true/false negatives were important to consider, and their methods for identifying negatives fell into three categories. 
First, and foremost, they are found in practice, and a few mentioned during manual investigations, and many mentioned through corroboration of multiple tools. 
A specific example given was the host AV finding a true positive that a network-level signature should have but did not find. 
Second, a respondent mentioned that anomaly detection systems are needed for exactly this, which supports the results of Figure \ref{fig:old-new-mc-tools} suggesting anomaly detection tools are becoming adopted. 
Third and most interestingly, a few analysts discussed using IOCs (indicators of compromise), which are intelligence shared among ally organizations on emerging attacks and their behaviors or mitigations. 
IOCs give or lead to new signatures, which multiple analysts said facilitate historical forensic efforts by running the new detectors against historic data  to find false negatives. 
One analyst discussed a more in-depth process including creation of a dashboard and using new intelligence and signatures for both automatic and manual hunting of threat in the previous 90 days of data. 
As before all three of these processes lead to re-configuration of tools and crafting of novel signatures.

{\it Questions 19, 20: What are strategies to reduce false positives and false negatives? Or to reduce the cost/time associated with false positives/negatives (e.g., how long it took to reconfigure/fix issues with a false positive)?: }
By far the most frequent response to this question was to manually reconfigure tools, in particular tuning rules/signatures. 
Drivers cited for tweaking tools included internal discoveries (e.g., false positives), new IOCs, user feedback, and conditioning rules per subnetwork. 
Interestingly, more automation for manually tuning rules was requested. 
Removing problematic tools was given as a response by two citing inaccuracies or continually needing reconfiguration as another. 
A general strategy of only adopting tools ``with a high signal-to-noise ratio'' was expressed. 
Prioritization was a common strategy including checking if a system is vulnerable as a first step of an investigation, using netflow analysis to isolate problem machines, and providing only summarized data to analysts for triage before more in-depth investigation. 
Other responses including white-/black-listing, writing plug-ins to fix problems, programatically fixing a tool by working with the vendor, and ``resolving (not fixing) issues, e.g., deleting problematic files.'' 

{\it Question 21, 22: How do you measure cost or efficiency of detection and prevention capabilities and novel tools in terms of the user's workstation? What is the current cost of security tools on the user's workstation in terms of memory requirements, processing time, user experience, or other metrics?: }
Six of the 13 respondents were unsure or said this was not applicable although two mentioned that this was taken into account by management in some opaque cost-benefit analysis. 
One analysts remarked that this is not taken into account as their security measures are non-negotiable, mandatory for compliance. 
Three mentioned performance measurements when testing on agents, specifically CPU time and memory used. 
In particular, some host logging tools also report metrics on their CPU and memory costs. 
Two mentioned disk I/O  as a metric, and perhaps the most important for actual use.  
To quote one respondent, 
\begin{quote}
    CPU and memory are not as important as disk I/O. Spinning disks cannot handle the host-based data collection client software. Reboot can be 15 minutes and involve too much performance costs. All that use SSD (steady state disks) can now accommodate the tools. 
    
    - Participant 9
\end{quote}
He/she went on to say that older Windows OSes are often to blame for user complaints of  performance  (not  necessarily  security  clients).   
Specifically, the ``enormous conglomeration of patches'' causes disk I/O  problems,  although  newer  versions  do  not  suffer  from this problem. 
This illustrates the familiarity of problems for the endpoint user that some analysts have necessarily developed.

No respondent gave quantifiable answers to the latter question of what the actual cost is in terms of the metrics they defined. 
Responses of the actual cost varied from indicating it was low, e.g. ``unsure, but I don't notice our clients on my host'', to it could be an issue depending on use with a potential to crash machines or cause disk failures. 
Multiple comments indicated that configuration of tools take into account host resources, specifically by scheduling taxing events such as scans or patching for non-work hours, or changing priorities of the tools' processes.

{\it Questions 23, 24: How do you measure cost or efficiency of detection and prevention capabilities and novel tools in terms of the analyst's time? What is the current cost of security tools on the analyst?: }
Actual dollar costs were common answers. 
One analysts gave an interesting formula for tool evaluation:  convert to dollars the human time spent on investigating false positives and sum this for all false positives from a given tool. 
Others cited actual figures, e.g., ``enforcing endpoint encryption was roughly a \$10M endeavor.'' 

Unsurprisingly the most common response to this question was simply quantification of analysts' time, i.e., the number of man hours or employees focusing on alert triage and response. 
The range of values cited in this category were between five and nine analysts full time focusing on alerts.
Emphasis is put on creating good workflows to reduce time per incident. 

Other similar metrics were mentioned including number of alerts, time per alert, number of total tickets, time per ticket, output of each analyst. 
Many were quick to note that while these fairly simple metrics were used they are insufficient and miss the point. 
One analysts remarked that counts of incidents or true positives or rates of dealing with alerts were perhaps unintuitively a red herring,  citing better tools may result in alerts that require more time or produce fewer false positives  (therefore getting worse scores in some simple metrics but providing greater security).  
Similar sentiment was expressed by others saying, 
\begin{quote} 
    The goal is empowering the analyst, not measuring time per ticket.
    
    - Participant 5
\end{quote}
and 
\begin{quote}
    We focus on ease of use and effectiveness of tools. 
    
    - Participant 10
\end{quote}
and 
\begin{quote}
    Time is less important than correct results. 
    
    - Participant 1
\end{quote}
Consequently, measuring security becomes much more nebulous. 
As one analyst answered, ``the amount of candy an analyst eats per day.'' 

In summary,  analysts quickly spouted off a variety of metrics to evaluate security, followed by backpedaling to make clear that these metrics do not provide true measurements of their understanding of an incident nor the effectiveness of the tools. 
This parallels what was found by Sundaramurthy et al. \cite{sundaramurthy2016turning}. To quote the authors, 
\begin{quote}
    The professional logic of security analysts (subject) dictates that they
constantly improve their skills and be efficient in detecting
and mitigating security threats. On the other hand, SOCs
are under constant pressure to demonstrate their value to
the parent organization. This results in a number of metrics
being defined to measure the performance of SOC analysts.
Ultimately, the job of the analysts is skewed very
much towards generating those defined metrics. This creates
a conflict within them. They are confounded with
two non-identical objectives---doing the right thing
versus the required thing.
\end{quote}

\section{Observed Themes \& Interesting Tangentials}
\label{sec:tangents}
Here we articulate broader takeaways from the responses that cross-cut the questionnaire sections. 
A benefit of an in-person/phone interview as opposed to a written survey or multiple choice questionnaire is the propensity for participants to describe interesting or impactful events that are tangential to the question at hand. 
While the above sections summarized targeted responses of each question, this section provides worthwhile discourse that did not fit into the reporting above with the goal of articulating worthwhile themes. 

{\it Handling and Using Data: }
Quantity of default-collected host data varies widely among organizations, but large (e.g., 100GB/day or more) is not uncommon. 
From an operation without the resources to collect widespread host logs, an analyst desired greater collections, while analysts with more comprehensive host data collections desired tools for understanding. 
We surmise that analysts believe having ample host data is better than none, but given ample data find it is hard to convert into usable information. 
Constraints on the quantity and usefulness of data are imposed by  cost in terms of hardware, indexing platforms, and analysts' time to handle/use the data effectively. 

Software for indexing and storing cyber data is seemingly a universal requirement, but cost is a limiting factor. 
Open-source solutions and less-costly commercial solutions are also being adopted, and many organizations have historic log storage.  
While this questionnaire did not inquire directly about how historic host logs are used, hunting threats with new intel to find previous false negatives was discussed by multiple participants. 

A common problem borne out of the responses was a general lack of event logging capabilities for Unix-based systems versus Windows and a lack of uniformity across logging data from different OSes. 
Another theme was the general availability of host data that was not default-collected but was retrievable by the SOC upon command. 
Specific examples included the hard drive (HD) images, memory dumps, individual files. 
This facilitates in-depth analysis (if data is retrieved soon after an event) without the burdensome collection of high-fidelity data from all hosts, but does not permit post-incident analysis if host data is not retrieved before it is lost (e.g., memory dumps).

{\it Less Data More Information, Please: }
A large and important theme  (also discussed in one form or another by previous studies \cite{goodall2004work, fink2006bridging}) was that ample data collection is accompanied by the burden of creating information and actionable knowledge from it, a currently manual process.  
The responses continually revisited the need for data collection capabilities that would  correlate heterogeneous sources of data to increase the signal and decreasing the noise.
To quote an analyst from our study, 
\begin{quote}
    Lots of work has gone into building the collection [capabilities]. More of the need is correlation and use of the data [for] faster and more efficient response.
    
    - Participant 12
\end{quote}

The lack of higher-level actionable information, when working with data associated with low-level events, was a common thread in the responses.  It appears that a significant interpretation gap exists; data at the singular event level is not necessarily sufficient for human understanding or manual response.  This was best expressed in the interviews by a participant who identified a challenge for analysts as,
\begin{quote}
Turning low-level system events into higher-level events and user-level actions

- Participant 5
\end{quote}
The state-of-the-practice approach to this information extraction leans heavily on human analysis.  In the interviews, the participants used terms like `integration', `correlation', and `anomaly detection' to express the need for tools and methods that analyze across the various low-level data, such that it is quickly interpreted and acted on by a human. 

Yet another example was given by a participant discussing vulnerability scanning tools. 
\begin{quote}
    [Vulnerability scanning \& management tool A] was supposed to fix the [Vulnerability scanning \& management tool B] problem---it gives a bunch of vulnerabilities but not an actionable item. What do we need to patch? [Vulnerability scanning \& management tool A] isolates the software product that is causing the vulnerabilities. It correlates vulnerabilities up the software level so we know what to patch. But we later found that [Vulnerability scanning \& management tool A] did not have a wide enough vulnerability database to compete with [Vulnerability scanning \& management tool B].
    
    - Participant 9
\end{quote}
One analyst digressed to describe a sequence of large-impact attacks against his organization spawning management mandates for increasing security posture and substantial increase funding and tools for their SOC.
He/she said that even with the increased funding and tools, the manpower and expertise needed for adequate forensic and log analysis is too high,
\begin{quote}
    Generally, we need log interrogation and host-based IDSes that are easier to use and understand than we currently have. ... The level of expertise needed for forensics and log analysis is high. This requires a lot of time and people power. Things like SysMon [that produce] a higher signature-to-noise ratio make thing understandable---off-the-shelf tools for filtering non-useful data. We'd like a capability that reduces the volume and focuses on high-fidelity logs. 
    
    - Participant 2
\end{quote}
Participant 2 went on to say later in the interview,
\begin{quote} 
Currently cyber operations are like a roach motel---logs go in but nothing comes out.
\end{quote} 
and that in addition to more analysts and more expertise, tools that increase the signal-to-noise ratio will help.

Our takeaway is that analysts foresee next generation tools in a two step process that (1) filter based on rules or other analytics to isolate a subset of security-relevant logs and (2) integrate related logs, not just within a host (e.g. process and creator process) but across heterogeneous streams of data (e.g., the shell commands, the socket, and packets of the spawned connection along with platform and user information for the host and only the relevant known vulnerabilities.)
We note that many efforts have taken on the research and development task to pioneer tools in this vein;
e.g., HoNe \cite{fink2005visual}, CLIQUE \cite{best2010real},   Stucco \cite{stucco}, SANSR \cite{huffer2017sansr} and IDEAS \cite{bridges2018ideas} propose visual-analytic tools to this end, while work of Hossain et al. \cite{hossain18dependence} and ProTracer \cite{ma2016protracer} seek to compress logs without inhibiting log analysis. 

{\it Analyst Time \& Effective Results: }
These cries for meaningful integration of data sources to help the data-to-decision process lead the another cross-cutting theme, that the real goal is protection and correct understanding from the data, but to achieve this analysts do not have enough time. 
One participant remarked that in his/her organization analysts are given only ten minutes for triaging an alert because of the volume. 
Hence, volume and accuracy of alerts are very consequential. 
As evidenced in the Evaluation of Tools section (Section \ref{sec:eval}) the analysts' time/tool efficiency and was a primary driver for tool selection, but evaluation metrics were often considered insignificant because correct results and effectiveness of tools is the real goal. 
Another respondent digressed, 
\begin{quote}
    [Our organization] has a standard risk level (low, medium, high), and there is a protocol for each level. Everyone is following rules, doing what they are told to do to meet the letter of the law for minimum-security standards. 
    ...
    We don't have enough people who really understand this stuff. ``What products?'' is inconsequential if the users lack expertise. 
    Instead [of building expertise] we've bought commercial-off-the-shelf tools to protect ourselves. 
    
    - Participant 2
\end{quote}
We note that research of Sundaramurthy et al. \cite{sundaramurthy2016turning} found that the propensity for managerial decisions drove purchase of security tools to be obtain best security practices, as opposed to efficiency, resulting in a contradictory position for analysts, specifically use of the tool is orthogonal to SOC workflows.  
Furthermore, Sundaramurthy et al.  also found that standard operating procedures can be problematic because the job necessarily requires creativity during investigations. 
To quote an analyst from Sundaramurthy et al.'s study, 
\begin{quote}
    The procedure were turning us into robots. The procedures were so detailed at some point that all the analysts were doing was to click and fill in data. 
\end{quote}
Balancing SOC resources in terms of new tools, new procedures, flexibility versus structure,  analyst man hours, and analysts training is an inherent challenge, and trust in many tools is suspect by the professionals.

{\it Interactive Tools and Embedding Situational Expertise:}
Two of the interview questions (addressing analyst challenges, and controlling false positives/negatives) provided good insight into how analysts interact with the tools.  The discussion above highlighted the need for actionable information from the tools versus just the raw data, and identified a gap in user understanding when working with singular events.  People desire to interact with the tools at a decision-making level, where actions can be determined from higher-order data correlations.  

Based on the analyst responses, users are also forced to interact at a very low-level when configuring the tools, not just  when they are trying to interpret events. Although not explicitly expressed, a higher-order interaction when configuring the tools can be inferred as a need, based on the analyst descriptions of dealing with false positives and negatives.  In most cases, the respondents cited an iterative tinkering with rules, indicators of compromise, or configurations as the means to make their tools more effective in their local environment. For example,
\begin{quote}
    manual reconfiguration and tweaking rules, both internally and from outside intel (new threat info) 
    
    - Participant 11
\end{quote}

\begin{quote}
    Rules need to be as specific as possible, to reduce possibility of false positives, saves analyst time.  Specificity can include tailoring signatures to the organization, eg. by zone (dmz, etc.)
    
    - Participant 4
\end{quote}
    
\begin{quote}
    initial strategy with <a vendor>: tune it and continual reconfig 
    
    - Participant 2
\end{quote}

These experiences highlight the need for a more interactive and/or more automated means to tune the tools to the nuances of the local network environment.  Participant 12 explicitly described a need for automation in tuning rules.  However, automation implies an interaction with the tool at a very high level.  For example, the tool could provide a means for analysts to enter feedback on the validity of an alert, and have that feedback be the basis for automated tuning. Although a non-trivial engineering task for a vendor, this would allow local analysts to embed their domain expertise and local experience into the tool.  Not only would this greatly simplify the tuning process and free analysts to focus on discovering threats, it is a way to permanently capture the human knowledge of the variation and customization of a local network.

{\it Trending Tools: }
Building dashboards and visualizations seemed an attractive  and indeed used option by our respondents, and not just for default, daily monitoring, but also for infrequent projects such as visualizations of what network and host data they collect to make new tool purchase decisions, or visualizations for hunting threats in historical data with new IOCs. 

Discussions of dynamic capabilities, e.g., changing the quantity or quality of data collected in certain situations, have been adopted. 

Traditional signature-based endpoint protection (AV) is being replaced or supplemented by behavior or machine-learning-based detectors. 
To quote an interviewee reported by Botta et al. \cite{botta2007towards}, 
\begin{quote}
    I would like a tool that could watch trends over time...what's normal patterns for our network, what's not normal patterns.
\end{quote}
This pre-2007 desire seems to preempt the anomaly detection algorithms and AI-based AV that are becoming adopted currently. 

Given that storing 90 days of historical data is becoming commonplace, we predict that large-scale analysis of data, e.g., training a learning algorithm on all historic logs, are a likely avenue for novel, network-specific technologies. 
Looking even more into the future, cloud-based infrastructures with remote desktops have the potential to replace current infrastructure. 
For SOC operations this means an reversal of position---instead of complete access to network-level data and potentially limited host data access, host-level security measures will dominate and network-level will be opaque. 
Furthermore, costs of processing and moving data will be charged per memory, storage byte, etc. 
Consequently, developing tools that integrate host-based data to pare-down size but increase information-content will likely be an important avenue for future technologies.

\section{Conclusion}
\label{sec:conclusion}
Security operations now have widespread collection and query capabilities for a diverse set of logs as well as configurable tools for meeting network-wide security requirements (e.g., end-point encryption, email management, etc.). 
In order to provide network protection,  continual investigations, log analysis, and tool configuration are commonplace for security analysts, but the ever-growing quantity of tasks  and data is proving problematic. 
As researchers seek to develop useful security technologies, understanding the burden and desires of the security operators is a necessity, but is hindered by the the limited access of outsiders to SOC data, problems, and operators. 
To this end, we conducted and report results of a survey of security analysts that probes their current practices, problems, and desires  with respect to host-based data collection, tools, and evaluation of tools. 
Our results give details of quantity and types of data collected, trending tools, evaluation procedures and metrics, and perhaps most importantly desired capabilities and current problems. 
Most notably, the analysts time is severely constrained because the volume of alerts is high, the accuracy of alerts is suspect, and investigations are difficult and time-consuming. 
Converting data into actionable information requires expertise and manual integration of multiple data sources, for which there is a great need to decrease data volume, increase the data's information content, and increase the analysts. 
We hope that our findings can assist future efforts to build worthwhile tools to enhance security.

% Modern security operations centers (SOCs) employ a variety of tools for intrusion detection, prevention, and widespread log aggregation and analysis. 
% While research efforts are quickly proposing novel algorithms and technologies for cyber security,  access to actual security personnel, their data, and their problems are necessarily limited by security concerns and time constraints. 
% To help bridge the gap between researchers and security centers, this paper reports results of a survey of 13 professionals from five different security operation centers (SOCs) including at least one large academic,  research, and government organization. 
% The survey consisted of 26 questions probing the current practices and future desires of SOC operators about host-based data collection capabilities, what is learned from the data, what tools are used,  and how tools are evaluated. 
% Responses are organized and reported by topic, then broader themes are 
% Forest-level takeaways from the results center on problems stemming from size of data, correlation of heterogeneous but related data, signal-to-noise ratio of data, and analysts' time. 

\section*{Acknowledgments}
The authors also thank all security analysts that agreed to participate in this study, and those individuals who helped with arranging the interviews.  
We also wish to thank Kerry Long for fruitful conversations contributing to this work's scope and direction. 
The research is based upon work supported by the Office of the Director of National Intelligence (ODNI), Intelligence Advanced Research Projects Activity (IARPA), via the Department of Energy (DOE) under contract  D2017-170222007. 
The views and conclusions contained herein are those of the authors and should not be interpreted as necessarily representing the official policies or endorsements, either expressed or implied, of the ODNI, IARPA, or the U.S. Government. 
The U.S. Government is authorized to reproduce and distribute reprints for Governmental purposes notwithstanding any copyright annotation thereon.

\balance  %balance the two columns at the end of the paper
\small
\bibliographystyle{abbrv}
\bibliography{refs}  
\end{document}